\documentclass{elsarticle}
\journal{Journal of Computational Physics}

\usepackage[utf8]{inputenc}
\usepackage{graphicx}
\usepackage{amsmath,amssymb}
\usepackage{caption}
\usepackage{subcaption}
\usepackage{xcolor}
\usepackage{hyperref}
\usepackage{ulem}
\usepackage{makecell}
\usepackage{bm}
\usepackage[margin=1.2in]{geometry}
\newcommand{\dd}[2]{\frac{d #1}{d #2}}

\newcommand{\avg}[1]{\left\langle #1 \right\rangle}

\newcommand{\R}{\mathbb{R}}
\newcommand{\mat}[1]{\begin{bmatrix}#1\end{bmatrix}}

\newcommand{\mus}{{m_{us}}}

\newcommand{\integrate}{\int_{0}^{T}}

\newcommand\reout[1]{}
\newcommand\reeout[1]{}
\newcommand\naxout[1]{}

\begin{document}

\begin{frontmatter}
\title{Sensitivity analysis on chaotic dynamical systems by Finite Difference Non-Intrusive Least Squares Shadowing (FD-NILSS)}

\author[angxiu]{Angxiu Ni}\corref{cor}
\cortext[cor]{Corresponding author.}
\ead{niangxiu@gmail.com}
\ead[url]{https://math.berkeley.edu/~niangxiu/}
\address[angxiu]{Department of Mathematics, University of California, Berkeley, CA 94720, USA}

\author[mit]{Qiqi Wang}
\ead{qiqi@mit.edu}
\author[mit]{Pablo Fernández}
\ead{pablof@mit.edu}
\address[mit]{Department of Aeronautics and Astronautics, Massachusetts Institute of Technology, Cambridge, MA 02139, USA}

\author[Chai]{Chaitanya Talnikar}
\ead{chaitukca@gmail.com}
\address[Chai]{Nvidia Corporation, Santa Clara, CA 95051, USA}

\begin{abstract}
We present the Finite Difference Non-Intrusive Least Squares Shadowing (FD-NILSS) algorithm for computing sensitivities of long-time averaged quantities in chaotic dynamical systems. 
FD-NILSS does not require tangent solvers, and can be implemented with little modification to existing numerical simulation software.
We also give a formula for solving the least-squares problem in FD-NILSS, which can be applied in NILSS as well.
Finally, we apply FD-NILSS for sensitivity analysis of a chaotic flow over a 3-D cylinder at Reynolds number 525, where FD-NILSS computes accurate sensitivities and the computational cost is in the same order as the numerical simulation.
\end{abstract}

\end{frontmatter}

\section{Introduction}
Many important phenomena in science and engineering, 
such as turbulent flows \cite{Kolmogorov_turbulence} and some fluid-structure interactions \cite{Dowell1982}, are chaotic. 
In these systems, the objectives are often long-time averaged rather than instantaneous quantities.
Non-Intrusive Least Squares Shadowing (NILSS) is a method to compute sensitivities of long-time averaged objectives in chaotic dynamical systems.
In this paper, we present the finite difference NILSS (FD-NILSS), and apply it on a chaotic flow past a 3-D cylinder.

Sensitivities are derivatives of objectives: they can help scientists and engineers design products \cite{Jameson1988,Reuther}, 
control processes and systems \cite{Bewley2001,Bewley2001a}, solve inverse problems \cite{Tromp}, 
estimate simulation errors \cite{Becker2001,Giles2002,Fidkowski}, assimilate measurement data \cite{Thepaut1991,COURTIER1993} 
quantify uncertainties \cite{Marzouk2015}, 
and train neural networks \cite{deeplearning_book_Goodfellow,linearRange_GD}.
When a dynamical system is chaotic, computing meaningful sensitivities is challenging.
In fact, conventional sensitivity methods do not converge for chaotic systems.

Many attempts have been made to overcome issues encountered by conventional sensitivity analysis methods.
Ruelle proved a linear response formula for SRB measures \cite{Ruelle_diff_maps,Ruelle_diff_maps_erratum,Ruelle_diff_flow},
which was implemented in the ensemble method developed by Lea and others \cite{Lea2000,eyink2004ruelle},
however, computational cost for ensemble methods are high for large systems \cite{Chandramoorthy_ensemble_adjoint}. 
Ruelle also gave the fluctuation dissipation theorem for systems far from equilibrium \cite{Ruelle_newFDT}, 
which describes the evolution of SRB measures due to perturbations on the governing equation.
This method was implemented by Abramov and Majda \cite{abramov2007blended,Abramov2008},
and by Lucarini and others \cite{lucarini_linear_response_climate2,lucarini_linear_response_climate}.
Ruelle's fluctuation dissipation theorem is an overshoot for our problem, 
since we only care the final change of SRB measure but not its history of evolution.

The Least Squares Shadowing (LSS) method developed by Wang and others \cite{wang2014convergence,Wang_ODE_LSS} 
computes the sensitivity of long-time averaged objectives by first computing an approximation of the shadowing direction.
It has been proved that under ergodicity and uniform hyperbolicity assumptions, LSS computes correct sensitivities \cite{Chater_convergence_LSS}. 
LSS has been successfully applied to sensitivity analysis in chaotic 2-D flows over an airfoil \cite{Blonigan_lss_airfoil}.

By developing a `non-intrusive' formulation of the least squares problem in LSS, 
the Non-Intrusive Least Squares Shadowing (NILSS) method \cite{Ni_NILSS_AIAA_2016,Ni_NILSS_JCP} 
constrains the minimization to only the unstable subspace.
For many engineering problems, the dimension of the unstable subspace
is much lower than the dimension of the dynamical system, 
and NILSS can be thousands times faster than LSS.  
Moreover, the `non-intrusive' formulation allows NILSS be implemented with little modifications to existing tangent solvers.
NILSS has been applied on complicated problems such as 2-D flow over backward steps \cite{Ni_NILSS_JCP} 
and aero-elastic oscillation of a 2-D airfoil \cite{Ni_NILSS_AIAA_2016}.

Ni recently defined and proved the unique existence of the adjoint shadowing direction,
and showed that adjoint shadowing directions can be used for adjoint sensitivity analysis \cite{Ni_adjoint_shadowing}.
Based on this theoretical advancement, Ni and Talnikar developed the Non-Intrusive Least Squares Adjoint Shadowing (NILSAS) algorithm.
NILSAS does not require tangent solvers, and its computational cost is independent of the number of parameters.
NILSAS has been applied on a 3-D flow over a cylinder. \cite{Ni_nilsas}

We present the finite difference NILSS (FD-NILSS), where the tangent solutions in NILSS are approximated by finite differences, 
thus allowing the FD-NILSS be implemented with only a primal solver.
This enriches applications of FD-NILSS to engineering problems, since most numerical simulation software do not have accompanying tangent solvers.

This paper is divided into three parts.
First we review the NILSS algorithm.
Then we derive the FD-NILSS algorithm.
Finally, we apply FD-NILSS to a 3-D flow problem for sensitivity analysis.

\section{A review of Non-Intrusive Least Squares Shadowing (NILSS)}

\subsection{Preliminaries}

We consider a chaotic dynamical system with the governing equation:
\begin{equation} \label{e:dynamical_system}
  \dd{u}{t} = f(u,s), \quad u \rvert_{t=0} = u^0 + v^{*0} s + \sum_{j=1}^M w_{j}^0\phi_j \;.
\end{equation}
Here $v^{*0}, w_{j}^0\in\R^m$ are directions of potential perturbations to the initial condition $u^0$;
$s\in\R$ is the system parameter, which affects both the governing equation and the initial condition;
$\phi_j\in\R$ controls the perturbation on $u^0$ in the direction of $w^0_j$;
$f(u,s):\R^m\times \R \rightarrow\R^m$ is a smooth function.
We assume base parameters $s=\phi_j=0$, hence the base trajectory has initial condition $u^0$.
We call the ODE in equation~\eqref{e:dynamical_system} the primal equation, and its solution $u(t)$ the primal solution.
A numerical solver for equation~\eqref{e:dynamical_system} is called a primal solver.

The objective is a long-time average defined as:
\begin{equation} \label{e:average J}
  \avg{J}_\infty:= \lim\limits_{t\rightarrow\infty}\avg{J}_T, 
  \text{ where }\avg{J}_T:= \frac{1}{T}\integrate J(u,s)dt\;,
\end{equation}
and $J(u,s):\R^m\times\R\rightarrow\R$ is the instantaneous objective function.
We make the assumption of ergodicity \cite{walters2000introduction}, hence $ \avg{J}_\infty $ only depends on $s$.

For a finite trajectory on time span $[0,T]$, and for any $j$, 
if we make an infinitesimal perturbation on the initial condition through $\phi_j+\delta \phi_j$, 
the trajectory will be perturbed by $\delta u$, which satisfies:
\begin{equation}
  \dd{\delta u}{t} = \partial_u f \, \delta u, \quad \delta u \rvert_{t=0} = w_j^0 \delta\phi_j\;,
\end{equation}
where $\partial_u f \in \R^{m\times m}$ is the Jacobian matrix.
We define a time-dependent function $w_j(t):\R\rightarrow\R^m$ by:
\begin{equation} \label{e:define_homo_tangent}
  w_j = \delta u /\delta \phi_j \;,
\end{equation} 
then $w_j$ satisfies the following ODE with initial condition:
\begin{equation} \label{e:homo_tangent}
  \dd{w_j}{t} = \partial_u f \,  w_j , \quad w_j \rvert_{t=0} = w_j^0 \;. \
\end{equation}
$w_j$ reflects the perturbation in the trajectory due to perturbation in the initial condition.
We call $w_j$ a homogeneous tangent solution, and equation~\eqref{e:homo_tangent} a homogeneous tangent equation.

If we make an infinitesimal perturbation in the parameter $s+\delta s$, 
the trajectory will be perturbed by $\delta u$, which now satisfies:
\begin{equation}
  \dd{\delta u}{t} = \partial_u f \, \delta u + f_s \, \delta s, \quad  \delta u \rvert_{t=0} = v^{*0} \delta s \,.
\end{equation}
We define a time-dependent function $v^*(t):\R\rightarrow\R^m$ by:
\begin{equation} \label{e:define_inhomo_tangent}
  v^* = \delta u /\delta s \;,
\end{equation} 
then $v^*$ satisfies the following ODE with initial condition:
\begin{equation} \label{e:inhomo_tangent}
  \dd{v^*}{t} = \partial_u f \, v^* + f_s, \quad v^* \rvert_{t=0} = v^{*0} \;. \
\end{equation}
$v^*$ reflects the perturbation in the trajectory due to perturbation in the parameter $s$, 
which affects both the governing ODE and the initial condition in the direction of $v^{*0}$.
We call $v^*$ an inhomogeneous tangent solution, and equation~\eqref{e:inhomo_tangent} an inhomogeneous tangent equation.

A Characteristic Lyapunov Vector (CLV), $\zeta(t)$, is a homogeneous tangent solution whose norm behaves like an exponential function of time. 
That is, there are $C_1,C_2>0$ and $\lambda\in \R$, such that for any $t\in \R$,
\begin{equation}\label{e:CLV}
  C_1 e^{\lambda t }\|\zeta(0)\|  \le \|\zeta(t)\| \le C_2 e^{\lambda t }\| \zeta (0)\| ,
\end{equation}
where the norm is the Euclidean norm in $\R^m$, and $\lambda$ is defined as the Lyapunov Exponent (LE) corresponding to this CLV. 
CLVs with positive LEs are called unstable, CLVs with negative LEs are called stable, and with zero LEs are neutral.
In this paper, the $j$-th largest LE and its corresponding CLV will be referred as the $j$-th LE and $j$-th CLV, respectively.

We assume that our system has a bounded global attractor which is uniform hyperbolic.
A bounded global attractor is a bounded set of states such that no matter what initial condition the system starts from, 
the trajectory will eventually enter the attractor and never leave.
Furthermore, we assume that all trajectories on the attractor are representative in the long-time behavior of the chaotic system.
Uniform hyperbolicity requires that the tangent space, at all states on the attractor, can be split into stable subspace, unstable subspace,
and a neutral subspace of dimension one.
Under our assumptions, we can show that the angles between all CLVs are larger than a positive angle, 
regardless of where we are on the attractor.

\subsection{Sensitivity analysis via shadowing methods} \label{s:shadowing methods}

With the assumptions made in the last subsection, there exists for each trajectory a shadowing direction $v^\infty$, 
which is an inhomogeneous tangent solution of equation (\ref{e:inhomo_tangent}), and its orthogonal projection  perpendicular to the trajectory,
$v^{\infty \perp}$, is uniformly bounded on a infinitely long trajectory \cite{Chater_convergence_LSS}.
The orthogonal projection $p^\perp(t)$ of some vector valued function of time, $p(t)$, is:
\begin{equation}\label{e:vperp projection}
  p^\perp (t) = p (t) - \frac{f^T(t) p(t)}{f^T(t) f(t)} f(t) \; ,
\end{equation}
where $f$ is the trajectory direction as defined in equation (\ref{e:dynamical_system}), and $\cdot ^T$ is the matrix transpose.

\begin{figure}[htb]
  \centering
  \includegraphics[trim=0cm 0cm 0cm 0cm, clip=true, width=0.5\textwidth]{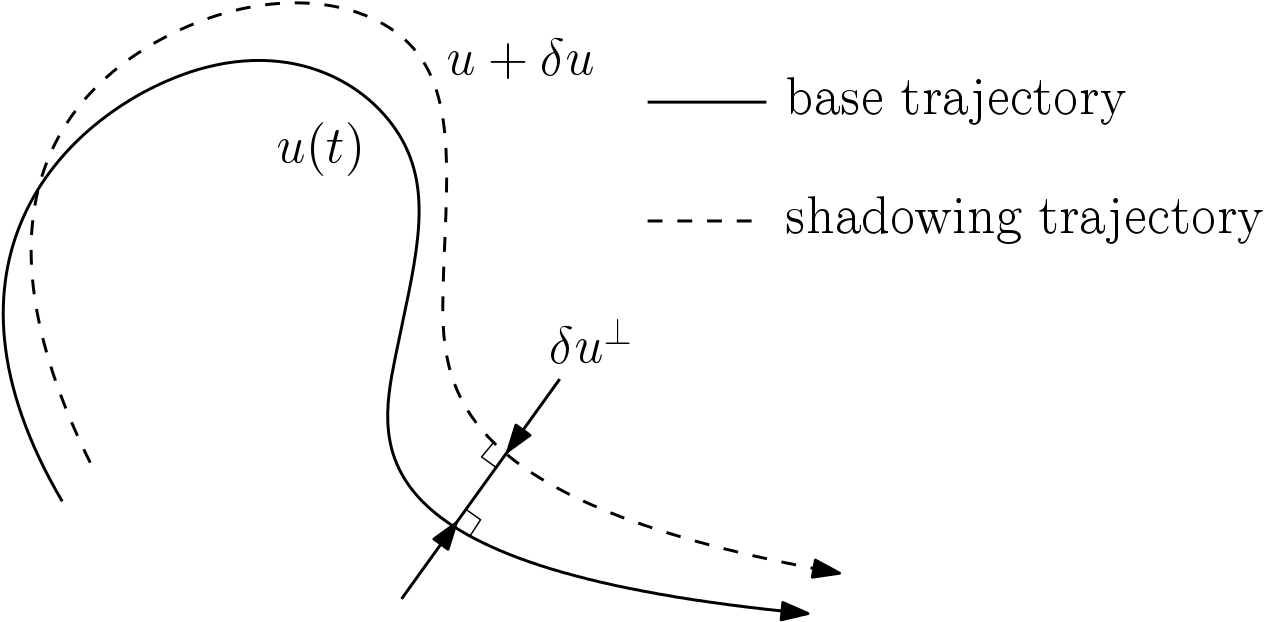}
  \caption{Example of a shadowing trajectory.
    The base trajectory has parameter $s$, the shadowing trajectory has parameter $s+\delta s$. 
    The first order approximation of $\delta u ^ \perp$ is $v^{\infty\perp} \delta s$, where $v^\infty$ is the shadowing direction.}
  \label{f:shadowing trajectory}
\end{figure}

The existence of shadowing directions means that,
there exists a new trajectory, defined as the shadowing trajectory, with perturbed parameter $s+\delta s$, 
such that $\delta u ^\perp$ is always smaller than $C\delta s$. 
Here $C$ is some constant, $\delta u$ is the difference between the new and the base trajectories,  
and $\delta u ^\perp$ is the perpendicular distance, as shown in figure \ref{f:shadowing trajectory}.
With the shadowing trajectory and the base trajectory close to each other, their difference, represented by the shadowing direction, 
can be used to compute the sensitivity $d\avg{J}_\infty/ds$.

For continuous dynamical systems, the base and the shadowing trajectories may move at different speed.
Since we are considering averages taken with respect to time, we should take account of the fact that if the shadowing trajectory
spend longer or shorter time in a particular neighborhood, 
then the weight of the objectives in this neighborhood should be respectively larger or smaller.
We define a `time dilation' terms $\eta$ to denote this effect, and if the shadowing trajectory takes less time to travel
the same length in the phase space, then $\eta <0$.
On the other hand, if the shadowing trajectory moves slower, then $\eta > 0$.
We can show that for a given inhomogeneous tangent solution $v$, which describes a perturbation on the trajectory due to parameter change, 
$\eta$ should satisfy the following equation \cite{Ni_NILSS_JCP}:
\begin{equation} 
  \dd{v^\perp}{t} = \partial_u f v^\perp + \partial_s f + \eta f\,.
\end{equation}
We denote the particular time dilation corresponding to the shadowing direction $v^\infty$ by $\eta^\infty$.

To conclude, the change in the location of the shadowing trajectory in the phase space is described by $v^\infty$;
the time difference the shadowing trajectory spend around a neighborhood is described by a corresponding $\eta^\infty$.
Taking both changes into account, we can can compute the sensitivity via:
\begin{equation} \label{e:djds with eta}
  \dd{\avg{J}_\infty}{s} \approx 
  \frac 1T \int_{0}^{T} \left[\partial_u J \, v^{\perp}+ \partial_sJ + \eta (J - \avg{J}_T) \right] \, dt \,,
\end{equation}
where we assume that $v$ and $\eta$ are legit approximations of $v^\infty$ and $\eta^\infty$.

Another formula for the sensitivity is easier for computer programming: 
\begin{equation} \label{e:derivative_1seg}
  \dd{\avg{J}_\infty}{s} \approx 
  \frac 1T \left[
  \int_{0}^{T} \left(\partial_u J \, v+ \partial_sJ \right)dt 
  +\xi \bigg\rvert ^T_0 \avg{J}_T
  -\left(\xi J \right) \bigg\rvert^T_0\right] ,
\end{equation} 
where the time difference term, $\xi$, is a time-dependent scalar function such that:
\begin{equation} \label{e:xi}
  \xi f = v-v^\perp \,.
\end{equation}
Intuitively, the right-hand-side of the above equation is how farther down the trajectory direction has the shadowing trajectory traveled.
Divided by $f$, we can see $\xi$ describes how much more time should the base trajectory take to catch up with the shadowing trajectory.
$\xi$ is easier to compute than $\eta$, since its definition does not involve time derivatives.
Notice that in equation~\eqref{e:derivative_1seg}, we use $v$ instead of its projection $v^\perp$.
The detailed derivation of equation~\eqref{e:djds with eta} and \eqref{e:derivative_1seg} can be found in the appendix of \citep{Ni_NILSS_JCP}.

Further, Wang et al. discovered that an approximation of $v^\infty$ can be found by minimizing the $L^2$ norm 
of inhomogeneous tangent solutions satisfying the ODE in equation~\eqref{e:inhomo_tangent},
since this minimization mimics the boundedness property of shadowing solutions.
The algorithm is call the Least-Squares Shadowing (LSS) method \cite{wang2014convergence,Wang_ODE_LSS,Chater_convergence_LSS}.
LSS has been successfully applied to several 2-D fluid problems, but the cost is typically high.

\subsection{The `non-intrusive' formulation} \label{s:NI formulation}

We briefly review the `non-intrusive' formulation of LSS and the NILSS algorithm developed by Ni et al. in \cite{Ni_NILSS_AIAA_2016,Ni_NILSS_JCP}.
Roughly speaking, we first represent the space of all inhomogeneous tangent equations
by a linear combination of all CLVs adding a particular inhomogeneous tangent solution.
Since unstable CLVs affect norms of the inhomogeneous tangent solutions the most, 
we can reduce the minimization in LSS to only the unstable subspace, 
which can be further approximated by the span of randomly initiated homogeneous tangent solutions.
One of the main technical difficulties following this idea is 
to separate the computation of the time dilation term from computing tangent solutions,
which is solved by introducing the perpendicular projection operator $\cdot^\perp$ we have used so far in this paper.
Another technical difficulty is to prevent over-growth of tangent solutions, which will be discussed in section \ref{s:divide to segments}.

More specifically, the NILSS problem on a given trajectory is a least squares problem with arguments $a \in \R^M$, 
where $M\ge \mus$, $\mus$ being the number of unstable CLVs:
\begin{equation} \label{e:nilss_1seg}
  \min_{a\in \R^M} \frac12\integrate ( v^{*\perp} + W^\perp a)^{T} ( v^{*\perp} + W^\perp a) \; dt ,
\end{equation}
Here $v^*$ is an inhomogeneous tangent solution, 
$W(t)$ is a matrix whose columns are orthogonal projections of randomly initialized homogeneous tangent solutions,
$W^\perp(t) = [w_1^\perp(t),\cdots,w_M^\perp(t)]$.
To conclude, the shadowing solution is given by $v = v^* + Wa$, which is an inhomogeneous tangent solution,
but we replace prescribing its initial condition by minimizing its $L^2$ norm.

We can further write equation~\eqref{e:nilss_1seg} into a formulation
where it is more obvious that this minimization is a least squares problem in $a$.
More specifically, equation~\eqref{e:nilss_1seg} is equivalent to
\begin{equation}
  \min_{a\in \R^M} \frac12 a^T \left[\integrate  (W^\perp)^T W^\perp \; dt \right] a
  + \left[\integrate  (v^{*\perp})^T W^\perp \; dt \right] a
\end{equation}
where we have neglected the constant term.
We denote the covariant matrix, the coefficient matrix for the second order terms, by $C$;
we denote the coefficient vector for the first order terms by $d$.
More specifically, we define
\begin{equation} \label{e:Cd_1seg}
  C = \integrate  (W^\perp)^T W^\perp \, dt \,,\quad
  d =\integrate  (v^{*\perp})^T W^\perp \, dt \,.
\end{equation}
Now the minimization problem is
\begin{equation} \label{e:nilss_1seg_aC}
  \min_{a\in \R^M} \frac12 a^T C a + d^T a \,.
\end{equation}

\section{Finite difference NILSS}

\subsection{Deriving the finite difference NILSS on a whole trajectory}

The FD-NILSS seeks to implement the NILSS algorithm with only primal numerical solvers,
which solves the primal equation in equation~\eqref{e:dynamical_system}.
This reduction on the requirement of accompanying solvers will make NILSS easier to implement and thus have more applications.
The main difficulty is primal solvers typically do not provide $W, v^*, \partial_u J, \partial_sJ$, and $\xi$ 
used in equation~\eqref{e:nilss_1seg} and \eqref{e:derivative_1seg}.
To resolve this difficulty, FD-NILSS approximately computes these unprovided quantities through finite differences.

Tangent solvers compute tangent solutions via solving the tangent equations,
whereas in FD-NILSS we compute tangent solutions via the their definitions in equations~\eqref{e:define_homo_tangent} and~\eqref{e:define_inhomo_tangent}.
More specifically, on a trajectory $u(t), t\in[0,T]$ with initial condition $u^0$,
to approximate a homogeneous solution $w_j$ with initial condition $w^{0}_j$, 
we compute primal solution $u^w_j$ by keeping the same $s$ but using initial conditions $u^0 + \Delta \phi_j w_j^0$, where $\Delta \phi_j$ is a small number.
The approximation for $w_j$ is thus
\begin{equation}\label{e:approx wj}
  w_j = \frac { \delta u }{ \delta \phi_j } \approx \frac{u^w_j- u}{\Delta \phi_j} \;.
\end{equation}
Since $W$ is a matrix whose columns are $\{w_j\}_{j=1}^M$, now with each column approximated via finite difference, 
we also obtain an approximation of $W$.

Similarly, to approximate an inhomogeneous tangent solution $v^*$ with initial condition $v^{*0}$, 
we compute primal solution $u^*$ with parameter $s+ \Delta s$ and initial condition $u^0 + \Delta s v^{*0}$.
The approximation for $v^*$ is thus
\begin{equation}
  v^* = \frac { \delta u }{ \delta s } \approx \frac{u^* - u}{\Delta s} \;.
\end{equation}
These approximations allows us to compute tangent solutions from primal solvers.
With those tangent solutions, we can compute $a$ via solving the minimization in equation~\eqref{e:nilss_1seg}, and compute the shadowing direction by $v = v^* + W a$.

We explain how to compute $\xi$ evaluated at $t=0,T$, which appear in equation~\eqref{e:derivative_1seg}.
At any $t$, the map $\psi:\R^m\rightarrow \R$ which maps $v(t)$ to $\xi(t)$ is a linear map defined as:
\begin{equation} \label{e:psiforxi}
  \psi(p) = \frac{p^Tf}{f^Tf} \;,
\end{equation}
where $p\in\R^m$, and $f$ is evaluated at $t$. 
Since we are expressing $v$ as $v=v^*+Wa$, we can compute $\xi$ from the same linear combination:
\begin{equation}
  \xi = \psi (v) = \psi(v^* + Wa) = \psi(v^*) + [\psi(w_1),\ldots,\psi(w_M)]a \;,
\end{equation}
where $v^*$ and $\{w_j\}_{j=1}^M$ are computed via finite difference.
This way of computing $\xi$ saves computer memory, since we no longer need to store vectors $v$ and $\{w_j\}_{j=1}^M$ at $t=0$ and $t=T$;
instead, we only need to store scalars $\psi(v^*),\psi(w_1),\ldots,\psi(w_M)$ evaluated at $t=0$ and $t=T$.

Finally we explain how to approximate, via finite differences, terms in equation~\eqref{e:derivative_1seg} involving $\partial_uJ$ and $\partial_sJ$,
which are typically not provided in numerical primal solvers.
More specifically,
\begin{equation} \label{e:d4} \begin{split}
&\int_{0}^{T} \left(\partial_u J \, v+ \partial_sJ \right)dt \\
= &\int_{0}^{T} \left[\partial_u J \, ( v^* + W a) + \partial_sJ \right]dt  \\
= &\int_{0}^{T} \left(\partial_u J \, v^*  + \partial_sJ \right)dt + 
\sum_{j=1}^{M} a_{j}\int_{0}^{T} \left(\partial_u J \,  w_{j} \right)dt \\
= &\int_{0}^{T} \left(\partial_u J \, \frac{\delta u}{\delta s}  + \partial_sJ \right)dt + 
\sum_{j=1}^{M} a_{j}\int_{0}^{T} \left(\partial_u J \,  \frac{\delta u}{\delta \phi_j} \right)dt \\
= & \int_{0}^{T} \frac 1{\delta s} \left(J(s+\delta s) - J(s) \right)dt + 
\sum_{j=1}^{M} a_{j}\int_{0}^{T} \frac 1 {\delta\phi_j} \left(J(\phi_j+\delta\phi_j) - J(\phi_j) \right)dt \\
\approx & \int_{0}^{T} \frac 1{\Delta s} \left(J(s+\Delta s) - J(s) \right)dt + 
\sum_{j=1}^{M} a_{j}\int_{0}^{T} \frac 1 {\Delta\phi_j} \left(J(\phi_j+\Delta\phi_j) - J(\phi_j) \right)dt \\
=& \tilde{J}^* + \sum_{j=1}^{M} a_{j} \tilde{J}_j^w
\;.
\end{split} \end{equation}
Here $\tilde{J}^*$ and $\tilde{J}_j^w$ are defined as:
\begin{equation}\begin{split}
  \tilde{J}^*  &= \frac 1{\Delta s} \int_{0}^{T}  \left(J(s+\Delta s) - J(s) \right)dt \\
  \tilde{J}_j^w &= \frac 1 {\Delta\phi}\int_{0}^{T}  \left(J(\phi_j+\Delta\phi_j) - J(\phi_j) \right)dt \;,
\end{split}\end{equation}
where $J(s+\Delta s)$ is short for $J(u(s+\Delta s, \phi_1,\ldots,\phi_M, t), s+\Delta s)$, that is,
the instantaneous objective evaluated from using parameter $s+\Delta s$, while all $\phi_j$'s are fixed as base values.
Similarly, $J(\phi_j+\Delta\phi_j)$ is short for  $J(u(s, \phi_1,\ldots, \phi_j+\Delta \phi_j,\ldots, \phi_M, t), s)$.

\subsection{Dividing trajectory into segments} \label{s:divide to segments}

There are two numerical issues when applying FD-NILSS on a whole trajectory with a large time length $T$.
The first issue, similar to NILSS, is that tangent solutions become dominated by the fasted growing CLV,
as a result, the minimization problem in equation~\eqref{e:nilss_1seg_aC} becomes ill-conditioned.
The second issue, unique to FD-NILSS, is that the perturbation on the trajectory falls out of the linear region, 
thus finite differences no longer approximate tangent solutions.
For FD-NILSS we use a similar technique as NILSS to solve these issues, 
that is, dividing the whole trajectory into multiple segments, and rescaling at interfaces.

\begin{figure}[htb]\begin{centering}
  \includegraphics[trim=0cm 0cm 0cm 0cm, clip=true, width = 0.5\textwidth]{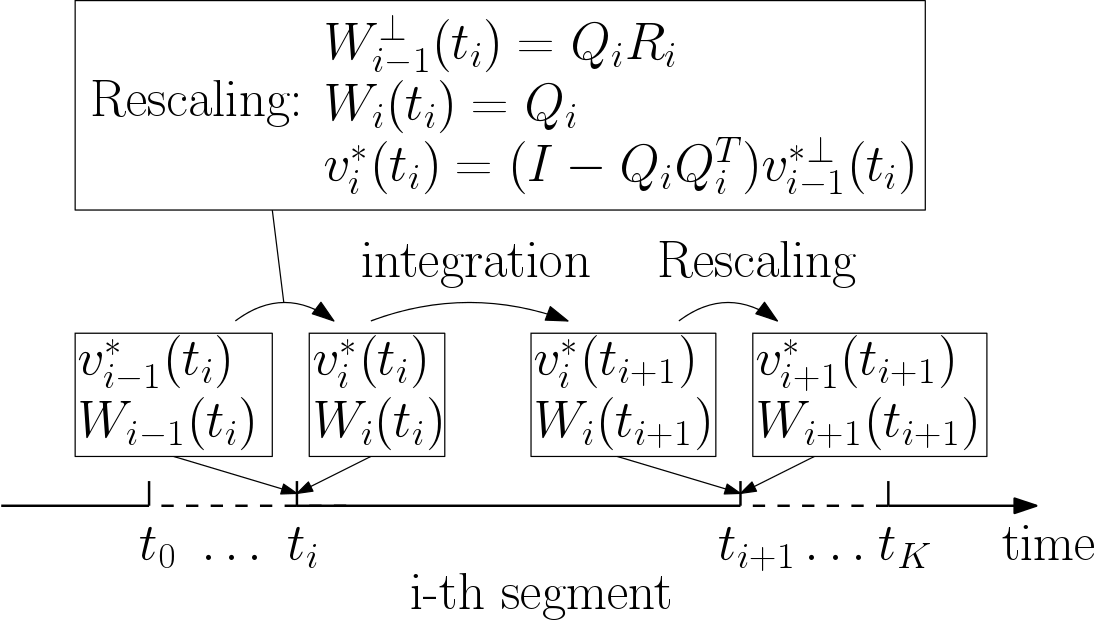}
  \caption{Subscripts used in this paper, where $t_0 = 0$, $t_K=T$.
  $W_i(t), v^*_i(t)$ are defined on the $i$-th segment, which spans $t\in[t_i, t_{i+1}]$; $Q_i, R_i$ are defined at $t_i$.
  Integrating tangent equations happens within one segment.
  Rescaling tangent solutions happens at the interface between two segments.
  }
  \label{f:subscript explain}
\end{centering}\end{figure}

We first describe how we select the subscript representing segment number for different quantities, as shown in figure~\ref{f:subscript explain}.
$T$ is the time length of the entire trajectory, which is further divided into $K$ segments, each of length $\Delta T$.
The $i$-th segment spans $[t_{i}, t_{i+1}]$, where $t_0=0, t_K = T$.
For quantities defined on a entire segment such as $W_i, v^*_i$, $C_i$, $d_i$ and $a_i$,
their subscripts are the same as the segment they are defined on.
For quantities defined only at the interfaces between segments such as $Q_i, R_i$, $b_i$ and $\lambda_i$,
their subscripts are the same as the time point they are defined at.

We use $W_{i-1}(t) = [w_{i-1,1}(t), \cdots w_{i-1,M}(t)]$, a $m\times M$ matrix-valued function of time,
to denote homogeneous tangent solutions on the $(i-1)$-th segment.
Assume that we have computed $W_{i-1}$, we explain how to generate initial conditions for $W_i$.
At time $t_i$, we first project all homogeneous tangent solutions to the subspace perpendicular to $f(t_i)$, to get $W_{i-1}^\perp(t_i)$,
upon which we then perform QR factorization, and use $Q$, 
the matrix with orthonormal columns, as the initial condition for $W_{i}$ on segment $i$.
More specifically, 
\begin{equation}\label{e:rescale_W}
  W_{i-1}^\perp(t_{i}) = Q_i R_i \,, \quad 
  \textnormal{and} \quad
  W_{i}(t_{i}) = Q_i \,.
\end{equation}

We use $v^*_{i-1}(t)$, a $m$-dimensional vector-valued function of time, 
to denote the particular inhomogeneous tangent solution on the $(i-1)$-th segment.
Assume that we have computed $v^*_{i-1}$, we explain how to generate initial conditions for $v^*_{i}$.
At time $t_i$, we first project to get $v^{*\perp}_{i-1}(t_i)$, 
from which we then subtract its orthogonal projection onto homogeneous tangent solutions.
More specifically, 
\begin{equation} \label{e:rescale_v*}
  b_i = Q_i ^T v^{*\perp}_{i-1}(t_{i}) \,,
  \quad
  \textnormal{and} \quad
  v^*_{i}(t_{i}) = v^{*\perp}_{i-1}(t_{i}) - Q_i b_i \,.
\end{equation}
This rescaling maintains the continuity of the affine space $v_i^{*\perp}+span\{w^\perp_{ij}\}_{j=1}^M$ across different segments.

The continuity of affine space allows us to impose continuity condition for $v_i^\perp$, 
which is the approximate shadowing direction on the $i$-th segment.
On each segment, define $v_i = v^{*}_i + W_i a_i$, where $a_i\in\R^M$.
The continuity condition can now be expressed via a relation between $a_i$ and $a_{i-1}$:
\begin{equation}
  v^{*\perp}_i(t_i) + W_i^\perp(t_i) a_i = v^{*\perp}_{i-1}(t_i) + W^\perp_{i-1}(t_i) a_{i-1} \;.
\end{equation}
Apply equation~\eqref{e:rescale_W} and~\eqref{e:rescale_v*}, and notice that $v^*_i(t_i)=v^{*\perp}_i(t_i)$, we get:
\begin{equation} 
  -Q_i b_i + Q_i a_{i} =  Q_i R_i a_{i-1}
\end{equation}
Since $Q_i$ has orthonormal columns, $Q_i^TQ_i=I\in \R^{M\times M}$. 
Multiplying $Q_i^T$ to the left of both sides, we have the continuity condition for $v$ at $t_i$:
\begin{equation}
  a_{i} = R_i a_{i-1} + b_i \;. 
\end{equation}

Due to the continuity condition, we can see that $v_i^\perp$ is continuous across segment.
However, $v_i$ is not continuous.
So there remains the question of how to construct a continuous $v$ and $\xi$ from $v_i$ and $\xi_i$, 
where $\xi_i$ is defined as such that $\xi_i(t) f(t)=v_i(t) - v_i^\perp(t)$,
so that we can apply equation~\eqref{e:derivative_1seg} to compute derivatives.
We give the following formula on the $i$-th segment:
\begin{equation} \begin{split}
  \xi(t) &= \xi_{i}(t) + \sum_{i'=0}^{i-1} \xi_{i'}(t_{i'+1}) \,, \\
  v(t) &= v_i^{\perp}(t) + \xi(t) f(t) \,. 
\end{split} \end{equation}
Readers may directly verify that above formula give continuous $v$ and $\xi$;
moreover, $v^\perp=v_i^\perp$ on all segments, $v$ is an inhomogeneous tangent solution, 
and $\xi$ and $v$ satisfy the pairing condition $\xi(t) f(t)=v(t) - v^\perp(t)$.

A further remark is that, we choose here to keep the continuity of the projected affine space $v_i^{*\perp}+span\{w^\perp_{ij}\}_{j=1}^M$.
We think it is also possible to choose to keep the continuity of the unprojected space $v_i^{*}+span\{w_{ij}\}_{j=1}^M$,
which should lead to easier derivation and programming.
However, we can not get rid of the projection process completely, since it still exists in the minimization step.
Moreover, this alternative approach would require to compute one more homogeneous tangent solution, 
since the neutral CLV is no longer projected out.
We suggest interested readers to try this possible approach.

\subsection{Procedure list of the FD-NILSS algorithm}

We should first prescribe the following parameters for FD-NILSS: 
\begin{itemize}
  \item Number of homogeneous tangents $M$, which should satisfy $M\ge \mus$, where $\mus$ is the number of unstable CLVs.
    Here we refer readers to \cite{Fernandez_LE_discretization} for how numerical discretization affects $\mus$, which in turn affects $M$.
  \item Perturbations $\Delta s, \Delta \phi_1, \ldots, \Delta \phi_M$. 
    Typically we set  $\Delta \phi_1 = \cdots = \Delta \phi_M = \Delta \phi$. 
    For convenience, we further set $\Delta s$ and $\Delta \phi$ to be the same small positive number $\epsilon$.
  \item length of each time segment $\Delta T$.
  \item number of time segments $K$.
\end{itemize}
Consequently, the time length of the entire trajectory, $T=K\Delta T$ is also determined.
The FD-NILSS algorithm is given by the following procedure list.

\begin{enumerate}

  \item Integrate equation (\ref{e:dynamical_system}) for sufficiently long time so that the trajectory lands onto the attractor. 
    Then, set $t = 0$.

  \item Generate initial conditions of homogeneous and inhomogeneous tangent solutions.
    
  \begin{enumerate}
    \item Generate a $m\times M$ random matrix $W^{0}= [ w_{1}^0, \cdots w_{M}^0 ]$.
      
    \item Compute  $W^{0\perp}= [ w_{1}^{0\perp}, \cdots w_{M}^{0\perp} ]$, whose column vectors are orthogonal to $f(t=0)$.

    \item \label{step:w0t0} Perform reduced QR factorization: $W^{0\perp} = Q_0 R_0$, where $Q_0 = [ q_{01}, \cdots q_{0M} ]$. 
      Since the span of columns in $Q_0$ is the same as that of $W^\perp_0$, columns in $Q_0$ are also orthogonal to $f(t=0)$.
      $Q_0$ will be the initial conditions for homogeneous tangent solutions.

    \item \label{step:v0t0} Set the initial condition for the inhomogeneous tangent solution: $v^*_0(t_0) = 0$.
  \end{enumerate}

  \item  For $i=0$ to $K-1$, on segment $i$, where $t\in [t_i, t_{i+1}]$ do:

  \begin{enumerate}
    \item Compute primal solutions and their related quantities. 

    \begin{enumerate}
      \item Compute the base trajectory $u(t)$ for $t\in [t_i, t_{i+1}]$ by integrating the primal system in equation~\eqref{e:dynamical_system}.

      \item Compute the instantaneous objective function $J(t)$ for the base trajectory. 

      \item Compute and store averaged objective on this segment, denoted by $\avg{J}_i$, and objective at the end of the segment $J(t_{i+1})$.
    \end{enumerate}

    \item Compute homogeneous tangent solutions and their related quantities. 

    \begin{enumerate}
      \item \label{step:compute uw}
        For each $1\le j \le M$, solve a solution $u^w_{ij}(t), t\in [t_i, t_{i+1}]$, of the primal system in equation~\eqref{e:dynamical_system}, 
        with initial condition $u^w_{ij}(t_i) = u(t_i) + \epsilon  q_{ij}$.
        Here $u(t_i)$ is the base trajectory at the beginning of segment $i$;
        $q_{ij}$ is given by step~\ref{step:w0t0} for the $0$-th segment and by step~\ref{step:w0ti} for later segments.

      \item The homogeneous tangent $w_{ij}(t)$ for $t \in [t_i, t_{i+1}]$ with initial condition $w_{ij}(t_i) = q_{ij}$ is approximated by: 
        \begin{equation}\label{e:approx w by f.d.}
          w_{ij} (t) \approx \frac{u^w_{ij} (t) - u (t)}{\epsilon} \,.
        \end{equation}
        Define an $m\times M$ matrix: $W_i(t) = [ w_{i1}(t), \cdots w_{iM}(t) ]$, $t\in[t_i,t_{i+1}]$.

      \item Compute orthogonal projection $W_i^\perp(t)= [ w_{i1}^\perp(t), \cdots w_{iM}^\perp(t) ]$ via:
        \begin{equation}\label{e:w projection}
          w_{ij}^\perp(t) = w_{ij}(t) - \frac{f^T(t) w_{ij}(t)}{f^T(t) f(t)} f(t) \; ,
        \end{equation}

      \item Compute and store the covariant matrix $C_i$ on segment $i$, defined as:
        \begin{equation} \label{e:integrate covariant matrix}
          C_i = \int_{t_{i}}^{t_{i+1}} (W_i^\perp)^T W_i^\perp dt. 
        \end{equation}
        This $C_i$ is the covariant matrix on the $i$-th segment, similar to that defined in equation~\eqref{e:Cd_1seg}.

      \item \label{step:w0ti} Perform reduced QR factorization: 
          $W_i^\perp(t_{i+1}) = Q_{i+1} R_{i+1}$, where $Q_{i+1}$ can be written in column vectors: $[q_{i+1,1}, \cdots q_{i+1,M}]$.

      \item For each $1\le j\le M$, compute and store $\xi^w_{ij}$:
        \begin{equation} \label{e:xiw_i}
          \xi_{ij}^w = \frac{( w_{ij}(t_{i+1}) )^T f(u(t_{i+1}))}{f(u(t_{i+1}))^T f(u(t_{i+1}))} \;.
        \end{equation} 
        Above term is $\psi(w_{ij})$ in equation~\eqref{e:psiforxi} evaluated at $t_{i+1}$.

      \item For each $0\le j \le M$, evaluate the instantaneous objective function on the trajectory with perturbed initial condition, $u_{ij}^w(t)$.
        We denote this perturbed objective function by $J^w_{ij}(t), t\in [t_i, t_{i+1}]$.
        Compute and store the perturbation in the time integration of the objective function:
        \begin{equation} \label{e:Jwi}
           \tilde{J}_{ij}^w = \frac 1\epsilon \int_{t_i}^{t_{i+1}} J^w_{ij}(t) - J(t) \, dt \,.
        \end{equation}
    \end{enumerate}

  \item Compute inhomogeneous tangent solutions and their related quantities.  
  \begin{enumerate}
    \item \label{step:compute u*}
      Solve a solution $u_i^*(t), t\in[t_i, t_{i+1}]$ of the primal system with parameter $s+\epsilon$,
      and initial condition $u^*_i(t_i) = u(t_i) + \epsilon v^*_{i}(t_i)$.
      Here $u(t)$ is the base trajectory;
      $v^*_{i}(t_i)$ is given by step~\ref{step:v0t0} for $0$-th segment and by step~\ref{step:v0ti} for later segments.

    \item The inhomogeneous tangent $v^*_i(t)$ for $t\in [t_i, t_{i+1}]$ with initial condition $v^*_i(t_{i})$ is approximated by:
      \begin{equation}\label{e:approx v* by f.d.}
      v^*_i\approx \frac{u^*_i-u}{\epsilon} \,.
      \end{equation}

    \item Compute the orthogonal projection $v_i^{*\perp}(t)$, $t\in [t_{i},t_{i+1}]$ via:
      \begin{equation}\label{e:v* projection}
        v_{i}^{*\perp} = v_{i}^* - \frac{f^T v_{i}^*}{f^T f} f \; ,
      \end{equation}

    \item 
      Compute and store
      \begin{equation} \label{e:integrate wTv*}
      d_i = \int_{t_{i}}^{t_{i+1}} {W_i^\perp}^T v^{*\perp}_i dt .
      \end{equation}
      This $d_i$ is similar to that defined in equation~\eqref{e:Cd_1seg}.

    \item \label{step:v0ti}
      Orthogonalize $v^{*\perp}_i(t_{i+1})$ with respect to $W^{\perp}_{i+1}(t_{i+1}) = Q_{i+1}$ 
      to obtain the initial condition of the next time segment:
      \begin{equation} 
        v^*_{i+1}(t_{i+1}) = v^{*\perp}_i(t_{i+1}) - Q_{i+1} b_{i+1} ,
      \end{equation}
      where $b_{i+1}$ is defined as:
      \begin{equation} \label{e:define_bi}
        b_{i+1} = Q_{i+1}^T v^{*\perp}_i(t_{i+1})\,, 
      \end{equation}
      and $b_{i+1}$ should be stored.

    \item Compute and store $\xi^*_i$:
      \begin{equation} \label{e:xi*_i}
        \xi_i^* = \frac{( v^*_i(t_{i+1}) )^T f(u(t_{i+1}))}{f(u(t_{i+1}))^T f(u(t_{i+1}))} \;.
      \end{equation}
      Above term is $\psi(v^*_i)$ in equation~\eqref{e:psiforxi} evaluated at $t_{i+1}$.

    \item Evaluate the instantaneous objective function on the perturbed trajectory $u_{i}^*(t)$.
      We denote this perturbed objective function by $J^*_{i}(t), t\in [t_i, t_{i+1}]$.
      Compute and store the perturbation in the time integration of the objective function:
      \begin{equation} \label{e:J*i}
        \tilde{J}_{i}^* = \frac 1\epsilon \int_{t_i}^{t_{i+1}} J^*_{i}(t) - J(t) \, dt \,.
      \end{equation}
  \end{enumerate}
  \end{enumerate}

  \item Solve the NILSS problem:
    \begin{equation} \label{e:NILSS_problem} \begin{split}
      &\min_{\{a_i\}} \sum_{i=0}^{K-1}  
      \frac 12 a_i^T C_i a_i + d_i^T a_i \\
      \mbox{s.t. }& 
      a_{i} = R_{i} a_{i-1} + b_{i}
      \quad i=1,\ldots,K-1.
    \end{split}\end{equation}
    This is a least-squares problem in $\{a_i\}_{i=0}^{i=K-1}$, where $a_i\in \R^M$ for each $i$.
    We give a suggestion on how to solve this least-squares problem in the next subsection.

  \item Compute 
    \begin{equation}\label{e:defineavgJT}
      \avg{J}_T = \frac 1K \sum_{i=0}^{K-1} \avg{J}_i \,.
    \end{equation}
    The derivative can be computed by:
    \begin{equation} \label{e:NILSS_sensitivity}
      \frac{d\avg{J}_\infty}{ds} \approx
      \frac 1T\sum_{i=0}^{K-1} \left[
        \tilde{J}^*_i 
        + \sum _{j=1}^M a_{ij} \tilde{J}_{ij}^w 
        + \left( \xi_i^* + \sum_{j=1}^M a_{ij} \xi_{ij}^w \right) \left(\avg{J}_T - J(t_{i+1})\right) \right] .
    \end{equation}
    Here $\tilde{J}^*_i$ is defined in equation~\eqref{e:J*i},
    $\tilde{J}_{ij}^w$ is defined in equation~\eqref{e:Jwi},
    $\xi_i^*$ is defined in equation~\eqref{e:xi*_i},
    $\xi_{ij}^w$ is defined in equation~\eqref{e:xiw_i}
    and $\avg{J}_T$ is defined in equation~\eqref{e:defineavgJT}.
\end{enumerate}
\qed

We first remark that there is no need to store $u_i$, $v^*_i$ or $W_i$ on the entire trajectory if we are only interested in the sensitivity.
The quantities that FD-NILSS needs are $C_i$, $d_i$, $R_i$, $b_i$ used in the minimization problem equation~\eqref{e:NILSS_problem},
and $\tilde{J}^*_i$, $\tilde{J}_{ij}^w$, $\xi_i^*$, $\xi_{ij}^w$, $J(t_{i+1})$ and $\avg{J}_T$ 
used in the sensitivity formula in equation~\eqref{e:NILSS_sensitivity}:
all of these quantities are either scalars, $M$-dimensional vectors or $M\times M$ matrices.
We should also store $u_i$, $v^*_i$ or $W_i$ at the end time of the last segment for resuming the algorithm or lengthening the trajectory.

The integrations in equation~\eqref{e:integrate covariant matrix}, \eqref{e:Jwi}, \eqref{e:integrate wTv*}, and \eqref{e:J*i}, 
can certainly be computed by summation over all time steps in the current time segment.
Alternatively, these integrations can be approximated by summation over several snapshots.
For example, the integration in equation~\eqref{e:integrate wTv*} can be approximated by:
\begin{equation}
  d_i\approx \frac{1}{2} \left( 
  W_i^{\perp T} v_i^{*\perp} (t_i) +
  W_i^{\perp T} v_i^{*\perp} (t_{i+1})  \right) \Delta T \;.
\end{equation}
Correspondingly, the finite difference approximations in equation~\eqref{e:approx w by f.d.} and~\eqref{e:approx v* by f.d.}
and the orthogonal projection in equation~\eqref{e:w projection} and~\eqref{e:v* projection},
now need be done only at the beginning and the end of a time segment.
Although taking snapshots does not reduce the computational complexity, it reduces data storage.
The idea of taking snapshots was also used in the multiple-shooting shadowing method developed by Blonigan \cite{Blonigan_MSS}.

The large part of the FD-NILSS algorithm is to compute $\{a_i\}_{i=0}^{K-1}$, 
using which we can construct the shadowing direction as shown in \cite{Ni_NILSS_JCP}:
this does not use any knowledge of the instantaneous objective function $J(u,s)$.
Hence the marginal cost for one more objective is almost negligible,
provided that we have determined all objectives before we run FD-NILSS.
If we stored all information generated during the computation, including all those primal solutions, 
then we may also add more objectives for very little cost, after the computation is done.

For one more parameter $s$, $\partial_s f$ is changed, hence $v^*$ is changed;
thus we need to recompute $\{a_i\}_{i=0}^{K-1}$, and the shadowing direction is also changed.
However, homogeneous tangents $W$ does not depend on $\partial_s f$,
hence the marginal cost for one more parameter in FD-NILSS is to compute another $v^*$,
and to solve again the NILSS problem in equation~\eqref{e:NILSS_problem}, 
whose cost is typically much lower than computing tangent solutions.
As a result, the marginal cost for one more parameter is about $1/M$ of the total cost,
provided that all parameters are determined before we run FD-NILSS.

\subsection{Solving the NILSS problem}

Here we give a suggestion on how to solve the minimization problem in equation~\eqref{e:NILSS_problem}.
The Lagrange function is:
\begin{equation}
  \sum_{i=0}^{K-1} \left(\frac 12 a_i^T C_i a_i + d_i^T a_i \right) + \sum_{i=1}^{K-1} \lambda_i^T \left( a_{i} - R_i a_{i-1} - b_i \right) \;.
\end{equation}
The Lagrange multiplier method tells us the minimizer for the NILSS problem is achieved at the solution of the following linear equation systems:
\begin{equation}\label{e:NILSS_block_matrix_form}
  \mat{ C & B^T\\ B & 0 }  
  \mat{ a\\  \lambda } 
  =\mat{ -d \\ b }\;,
\end{equation}
where the block matrices $C\in\R^{MK \times MK}$, $B\in \R^{(MK-M)\times MK}$,
vectors $a, d\in \R^{MK}$, and $\lambda, b \in \R^{MK-M}$.
More specifically, 
\begin{equation}\begin{split}
  & C = \mat{C_0 \\ &C_1 \\&& \ddots \\ &&&C_{K-1}}
  , \quad
  B = \mat{-R_1 & I \\ &-R_2 &I \\&&\ddots &\ddots \\ && &-R_{K-1}& I}, \\
  & a = \mat{a_0 \\ \vdots\\a_{K-1}}
  , \quad
  \lambda = \mat{\lambda_1 \\ \vdots\\ \lambda_{K-1} }
  , \quad
  d = \mat{d_0 \\ \vdots\\d_{K-1}}
  , \quad
  b = \mat{b_1 \\ \vdots\\b_{K-1} }
  ,
\end{split}\end{equation} 
where matrices $C_i, R_i \in \R^{M\times M}$, and vectors $a_i, \lambda_i, d_i, b_i \in \R^M$.

We can solve the Schur complement of equation~\eqref{e:NILSS_block_matrix_form} for $\lambda$:
\begin{equation}
  - B C^{-1} B^T \lambda = B C^{-1}d + b \;,
\end{equation}
where $C^{-1}$ can be computed via inverting each diagonal block in $C$.
Then we compute $a$ by:
\begin{equation} \label{e:expression_a}
  a = - C^{-1}(B^T \lambda +d) \;.
\end{equation}

The above formula for solving the least-squares problem in FD-NILSS can as well be used in NILSS \cite{Ni_NILSS_JCP}, which solves the same least-squares problem.
Moreover, if we use snapshots at the beginning of each time segment to replace the inner products between tangent solutions, 
then due to the orthonormalization procedures we have $C_i=I, d_i=0$, which further eases implementation.

\subsection{Remarks on FD-NILSS} \label{s:remarks}

We first notice readers that homogeneous tangent solutions computed in FD-NILSS can also be used to compute LEs.
More specifically, Benettin showed in \cite{Benettin1980_LE} that almost surely, in the long-time limit, 
the volume growth rate of the parallelepiped spanned by, say $M$, randomly initialized homogeneous tangent solutions, 
will be almost the same as the growth rate of the parallelepiped spanned by the first $M$ CLVs.
Now the $M+1$-th LE can be computed by subtracting the volume growth rate of the parallelepiped spanned by $M$ homogeneous tangent solutions
from the growth rate of the parallelepiped spanned by all previous $M$ plus one new homogeneous tangent solutions.

A caveat in Benettin's result is that when applied to a finitely long trajectory,
LEs may not show up in the exact descending order.
For example, if the random initial condition of the first homogeneous tangent solution 
happens to have only very small component in the direction of the first CLV,
then after finite time, we may still only observe the first tangent solution being dominated by the second CLV.
The same concern applies to NILSS and FD-NILSS, that is, we should typically compute some more homogeneous tangent solutions than exactly $\mus$,
in case the random initial conditions does not contain enough unstable components to cancel the exponential growth in $v^*$.
A sufficient $M$ for a particular set of initial conditions can be identified as that Benettin's algorithm has confidently given all positive LEs.

We required that $M$ be larger than $\mus$; however, we do not need to know $m_{us}$ or give $M$ a priori.
First, we can add tangent solutions to FD-NILSS inductively.
Assume that we currently have $M$ tangent solutions, then for equation~\eqref{e:NILSS_problem},
when adding one more tangent solution, then coefficients arrays $d_i$ and $b_i$ should each be augmented by one more entry, 
while the old coefficient arrays are not changed inside the new arrays;
$C_i$, $R_i$ should be augmented by one row and one column;
for equation~\eqref{e:NILSS_sensitivity}, we should further compute $\tilde{J}_{i,M+1}^w$ and $\xi_{i,M+1}^w$ for all $i$.
Second, due to the last comment, we can run Benettin's algorithm each time a new homogeneous tangent solution is added, 
and stop when all positive LEs has appeared, at which time we would have a big enough $M$.

We discuss how to select perturbation coefficient $\epsilon$ and segment length $\Delta T$.
These two algorithm parameters are mainly constrained by the requirement that finite differences adequately approximate tangent solutions.
Large $\epsilon$ would immediately yield a perturbed trajectory out of the linear approximation region;
on the other hand, too small an $\epsilon$ would lead to large computation error when subtracting the perturbed trajectory from the base trajectory.
Large $\Delta T$ would also allow the perturbed trajectory eventually falling out of the linear approximation region;
small $\Delta T$ would lead to high computational cost.
The solution is to run a linearity test.
More specifically, on the segment $[t_0, t_1]$, first select an $\epsilon$,
then compute $w_{01}$ by equation~\eqref{e:approx w by f.d.} from a random initial condition of unit length,
and compute $v^*_0$ by equation~\eqref{e:approx v* by f.d.} from a zero initial condition.
Then compute again using $2\epsilon$.
We select $\epsilon$ when the $w_{01}(t_1)$ and $v^*_0(t_1)$ computed using $\epsilon$ and $2\epsilon$ relatively differ less than some small number, say $\delta=0.01$.
Now the finite difference error leads to a $\delta$ relative error in $v$ and so hence in the sensitivity.
Notice we may need to adjust the unit of parameter $s$, or adjust $\Delta s$ and $\Delta \phi$ separately, 
if we can not find a common $\epsilon$ such that both $w_{01}$ and $v^*_0$ pass the linearity test.

The total time length $T$ is determined by the convergence history of sensitivity or by the computational cost requirement.
Typically, $T$ is determined empirically as the time when the sensitivity computed by FD-NILSS converges to within the uncertainty bound we desire.
Another possibility is to stop computation when the limited time or computation resource has passed.
We choose the latter criteria later in this paper, since we want to compare the cost of FD-NILSS to that of solving the primal system.
We have found that typically shadowing methods require a shorter trajectory to compute sensitivity than that required by the primal solver to reflect average behavior.

For problems with a large number of unstable Lyapunov exponents, we suggest switching from FD-NILSS to NILSS or NILSAS,
where we can take advantage of the vectorization of linear solvers and further accelerate computing homogeneous tangent or adjoint solutions,
as discussed in \cite{Ni_nilsas}.
Still, the cost of non-intrusive shadowing methods can get larger when $\mus$ is larger and the system becomes more chaotic.
However, such cost increase is typical for many numerical methods, for example, 
even computing long-time averages should take longer time to converge for more chaotic systems.
For very chaotic systems, if we still have $\mus \ll m$, then NILSS and NILSAS may still be competitive in computational efficiency;
at least, the idea of the `non-intrusive' formulation, that is, reducing the computation to unstable subspace, will still be important.
Current investigation on some computer-simulated fluid systems all have $\mus \le 0.1\% m$ 
\cite{Ni_NILSS_JCP,Fernandez_LE_discretization,Blonigan_channelLE,Ni_CLV_cylinder},
but we do not yet have a good estimation for very chaotic systems.
On the other hand, there are systems with $\mus \approx m$, such as Hamiltonian systems, which has equally many stable and unstable CLVs;
for these systems, NILSS or NILSAS may not be faster than other methods.

\section{Application on a turbulent three-dimensional flow over a cylinder}

\subsection{Physical problem and numerical simulation} \label{s:flow problem}

\begin{figure}[htb] \centering
  \includegraphics[width = 0.5\textwidth]{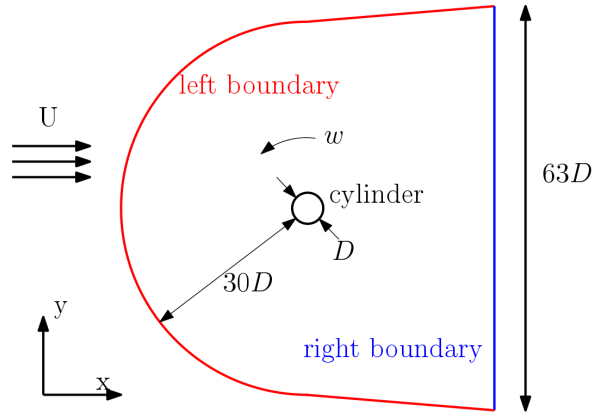}
  \caption{Geometry used in the simulation of a flow over a 3-D cylinder. 
    The span-wise extent of the computational domain is $Z=2D$. 
    The positive direction of the cylinder rotational speed $\omega$ is counter-clockwise.}
  \label{f:geometry}
\end{figure}

Before using FD-NILSS to compute sensitivities, we first describe the physical problem of the 3-D flow past a cylinder.
The front view of the geometry of the entire flow field is shown in figure \ref{f:geometry}.
The diameter of the cylinder is $D=0.25\times10^{-3}$.
The span-wise width is $Z=2D$.
The free-stream conditions are: density $\rho=1.18$, pressure $P = 1.01\times 10^5$, temperature $T=298$, dynamic viscosity $\mu=1.86\times10^{-5}$.
The free stream flow is in the x-direction, with the velocity $U$ being one of the system parameters, and for the base case $U_0=33.0$.
The flow-through time $t_0$, defined as the time for $U_0$ flowing past the cylinder, is $t_0 = D/U_0 = 7.576\times 10^{-6}$.
The Reynolds number of the base case is $Re=525$ and Mach number is $0.1$.
The cylinder can rotate around its center with rotational speed $\omega$, which is the second system parameter for our problem.
$\omega$ is measured in revolutions per unit time, and its positive direction is counter-clockwise, as shown in figure~\ref{f:geometry}.
For the cylinder to rotate one cycle per flow-through time, $\omega_0=1/t_0=1.32\times 10^5$.

\begin{figure}[htb]
  \centering
  \includegraphics[trim=3cm 0cm 7cm 0cm, clip=true, width=0.42\textwidth]{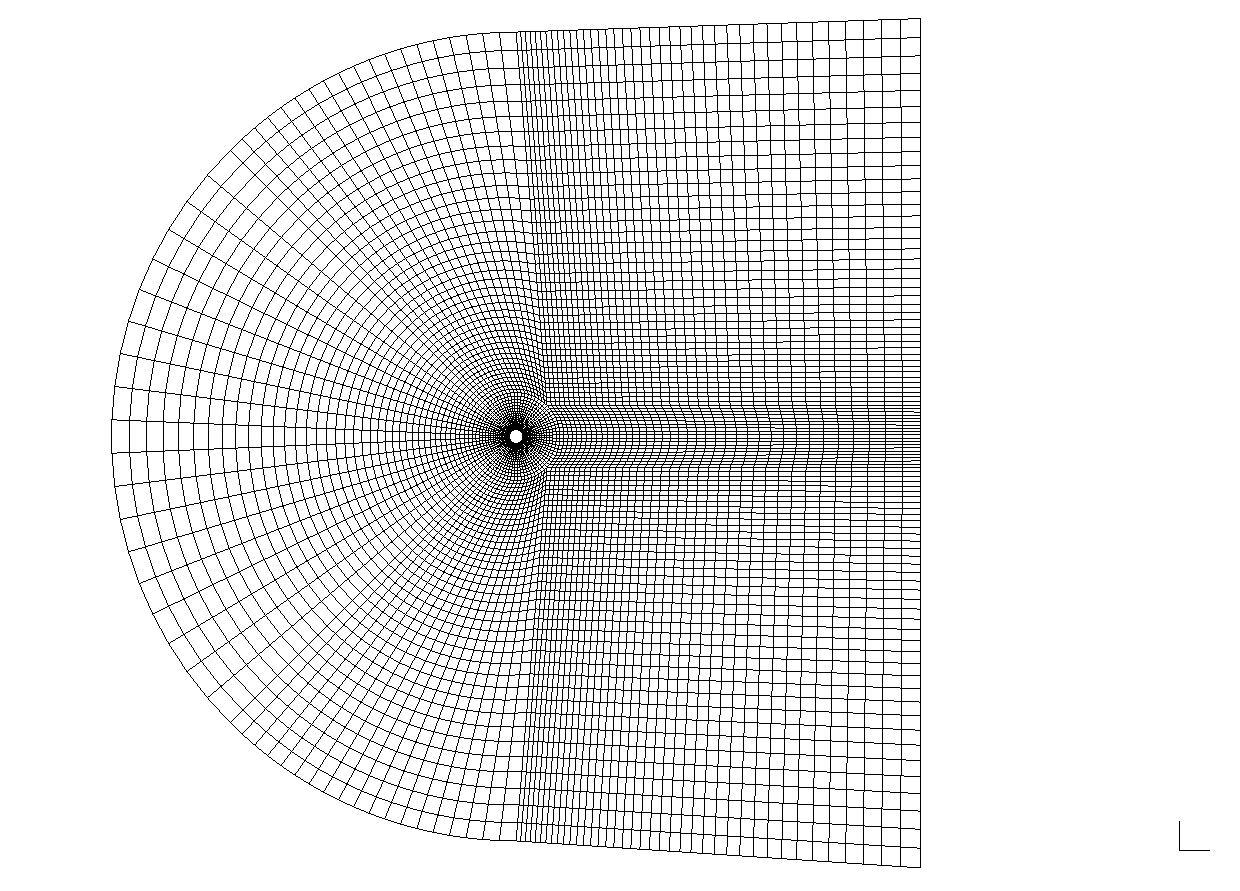}
  \includegraphics[trim=3cm 0cm 7cm 0cm, clip=true, width=0.42\textwidth]{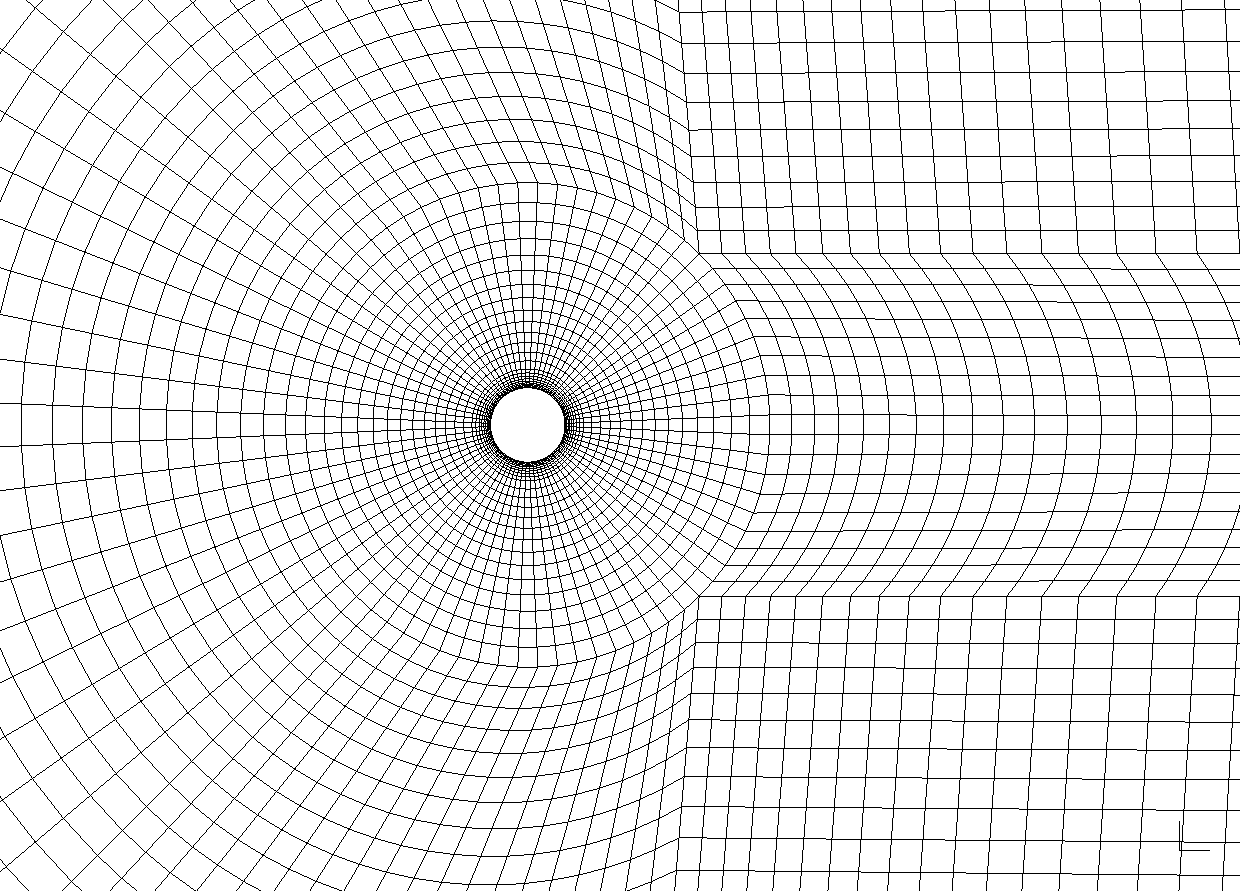}
  \caption{Left: 2-D slice of the mesh over the entire computational domain. 
    Right: zoom around the cylinder. 
    This is a block-structured mesh with $3.7\times10^5$ hexahedra.
    The span-wise direction has 48 cells.}
  \label{f:mesh}
\end{figure}

Then we look at settings for numerical simulations.
We use a block-structured mesh with $3.7\times10^5$ hexahedra.
2-D slices of the mesh are shown in figure \ref{f:mesh}.
The span-wise direction has 48 cells.
The CFD solver we use is CharLES developed at Cascade Technologies \cite{bres_Charles_solver}, 
using which we perform the implicit large eddy simulation, 
where the numerical error of the discretization scheme serves as the sub-grid scale Reynolds stress model.
The accuracy of the solver is formally 2nd order in space and 3rd order in time.
The span-wise boundary uses periodic boundary conditions; 
the left boundary uses a convective boundary condition \cite{Colonius1993};
the right boundary uses the Navier-Stokes characteristic boundary conditions (NSCBC) boundary condition \cite{Poinsot1992}.
The time step size is $\Delta t = 9.8\times10^{-9} = 1.30\times 10^{-3} t_0$.
In order to trigger the 3-D flow faster in our numerical simulation, 
we add a small white noise to the initial condition, whose magnitude is about $0.1\%$ of the inflow.

\begin{figure}[htb]
  \centering
  \includegraphics[trim=10cm 10cm 1cm 10cm, clip=true, width = 0.8\textwidth]{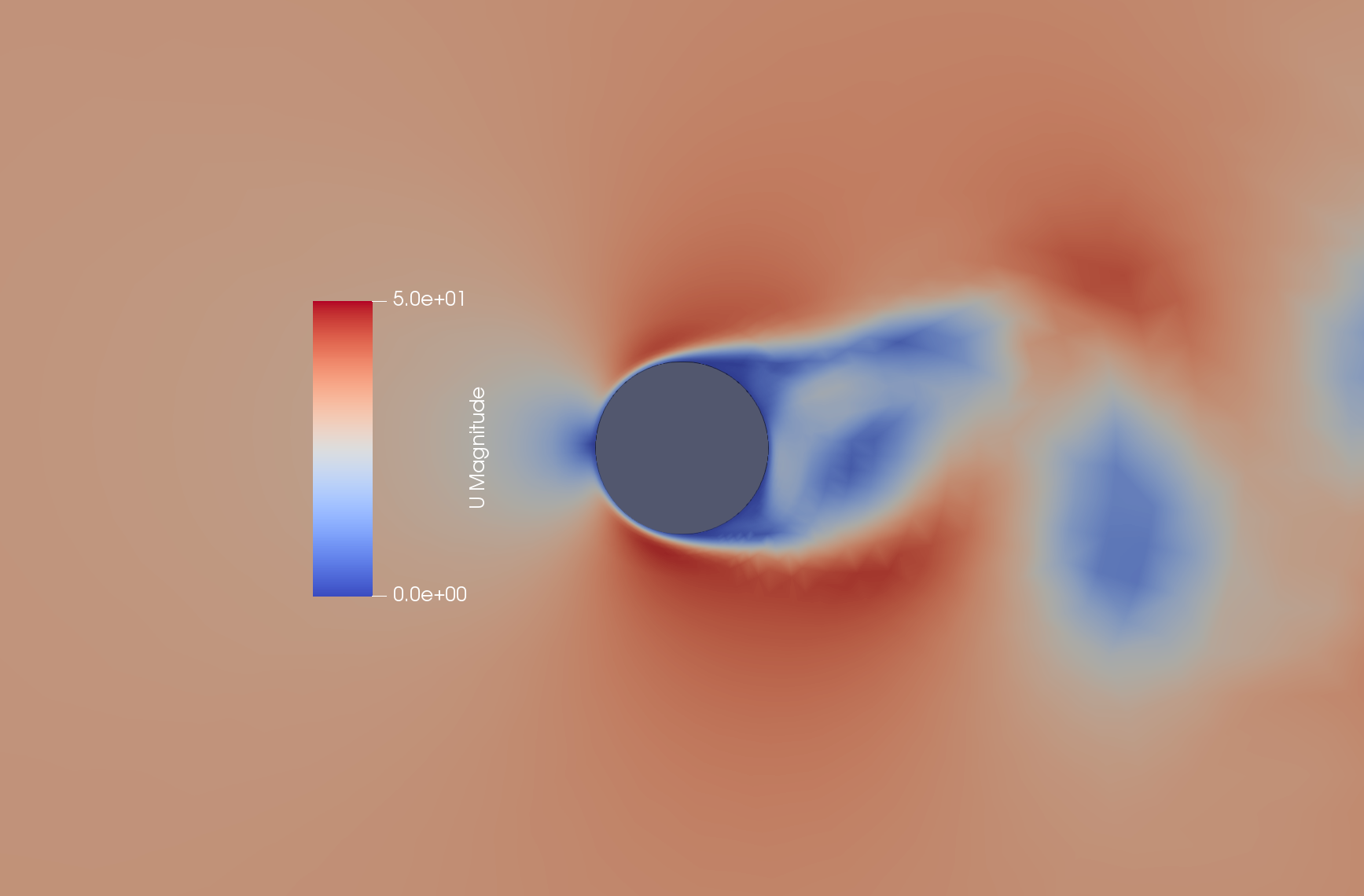}\\
  \includegraphics[trim=10cm 12cm 1cm 12cm, clip=true, width = 0.8\textwidth]{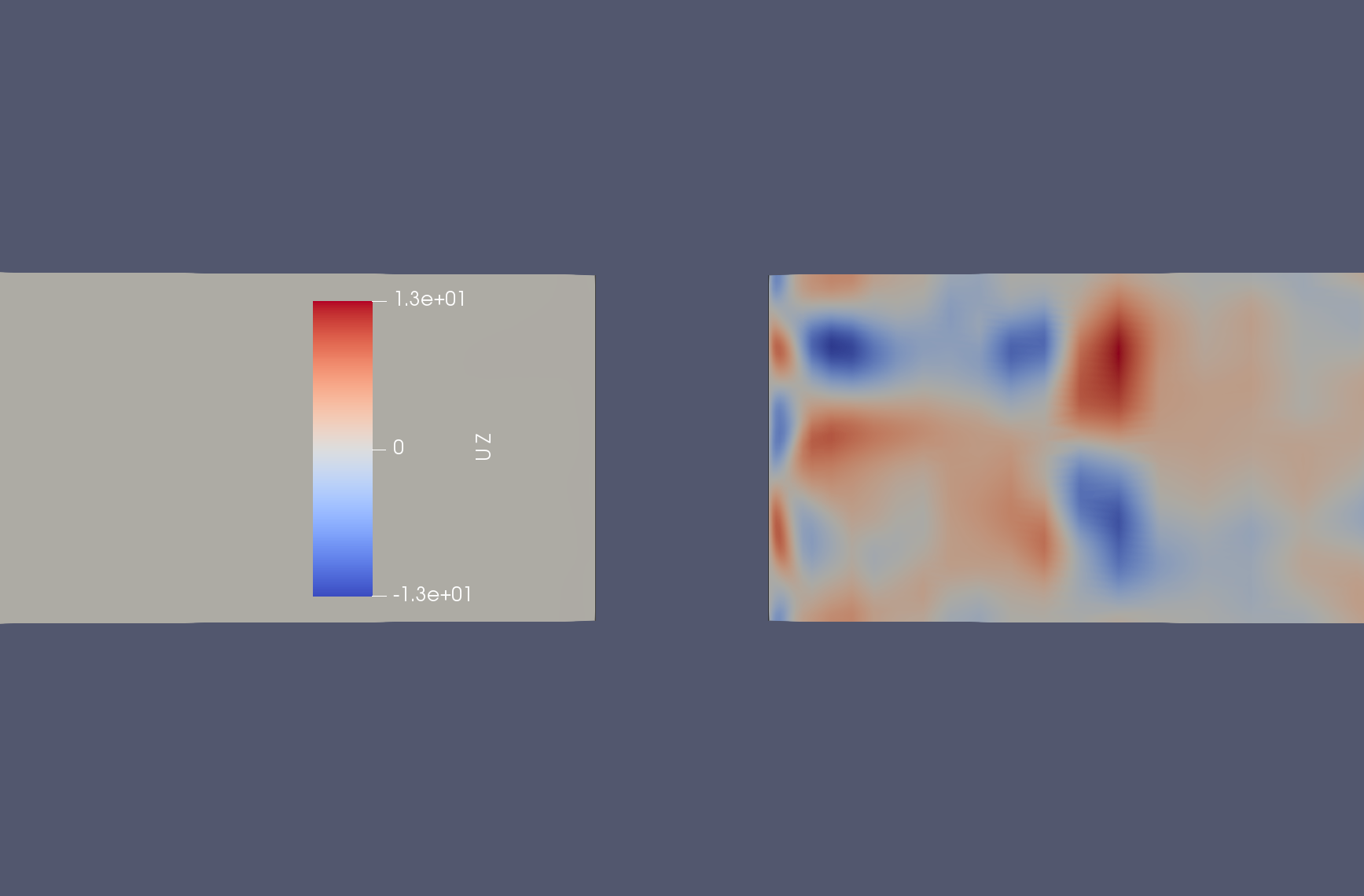}
  \caption{A typical snapshot of the flow field. 
      Top: cross-section along the x-z plane, plotted by magnitude of velocity.
      Bottom: cross-section along the x-y plane, plotted by the $z$-component of velocity. 
      The bottom picture shows the flow is 3-D.}
  \label{f:flow field}
\end{figure}

2-D snapshots of the flow field at $U=U_0$ are shown in figure \ref{f:flow field}.
The flow is chaotic and 3-D.
The same physical problem has been investigated through experiments by Williamson and Roshko \cite{williamson1990measurements}, 
and through numerical simulations by Mittal and Balachandar \cite{Mittal1996}.
The comparison of the Strouhal number and the averaged drag coefficient is shown in table \ref{t:verify}.
Here the Strouhal number is defined by $S_t=fD/U$, 
where $f$ is the main frequency of the vortex shedding, 
selected as the location of the peak in the Fourier transformation of the lift history;
the drag coefficient $C_D = D_r / (0.5\rho U^2 D Z)$, where $D_r$ is the drag.
As we can see, our simulation matches previous experimental and numerical results.

\begin{table}[htb]
  \setlength{\tabcolsep}{12pt}
  \centering
  \begin{tabular}{r l l }
    \hline
    & $S_t$ & $C_D$ \\ 
    \hline
    Current work 	                                    & 0.21  & 1.22  \\ 
    Previous 2-D simulation \cite{Mittal1996}               & 0.22  & 1.44  \\
    Previous 3-D simulation \cite{Mittal1996}	            & 0.22  & 1.24  \\
    Previous experiment \cite{williamson1990measurements}   & 0.21  & 1.15  \\
    \hline
  \end{tabular}
  \caption{Comparison of our simulation with previous results in literatures by the Strouhal number $S_t$ and the averaged drag coefficient $C_D$.}
  \label{t:verify}
\end{table}

\subsection{Results}\label{s:results}

We apply FD-NILSS to this 3-D chaotic flow past a cylinder.
\footnote{The python package `fds' implementing FD-NILSS is available at https://github.com/qiqi/fds. 
The particular files related to the application in this section are in fds/apps/charles\_cylinder3D.}
We consider two system parameters:
free-stream velocity $U$ and the rotational speed of the cylinder $\omega$. 
We will normalize $U$ by $U_0$, time by $t_0$, and $\omega$ by $\omega_0$.
We investigate the effect of $U$ on two objectives: averaged drag force $\avg{D_r}$, and averaged base suction pressure $\avg{S_b}$,
which is defined as the pressure drop at the base of the cylinder in comparison to the free stream.
We will normalize $\avg{D_r}$ by $F_0=0.5\rho U_0^2 D Z=8.031\times 10^{-5}$, and $\avg{S_b}$ by $P_0=0.5\rho U_0^2=642.5$.
For $\omega$, we look at its effect on averaged lift $\avg{L}$ and averaged lift square $\avg{L^2}$. 
We will normalize  $\avg{L}$ by $F_0$, and $\avg{L^2}$ by $F_0^2=6.450\times10^{-9}$.

\begin{figure}[htb]
  \centering
  \includegraphics[trim=0cm 0cm 0cm 0cm, clip=true, width=0.42\textwidth]{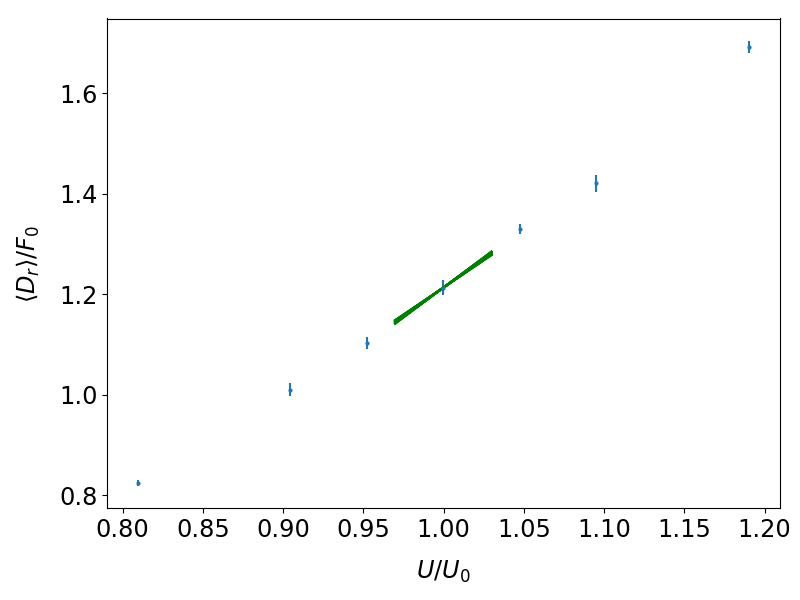}
  \includegraphics[trim=0cm 0cm 0cm 0cm, clip=true, width=0.42\textwidth]{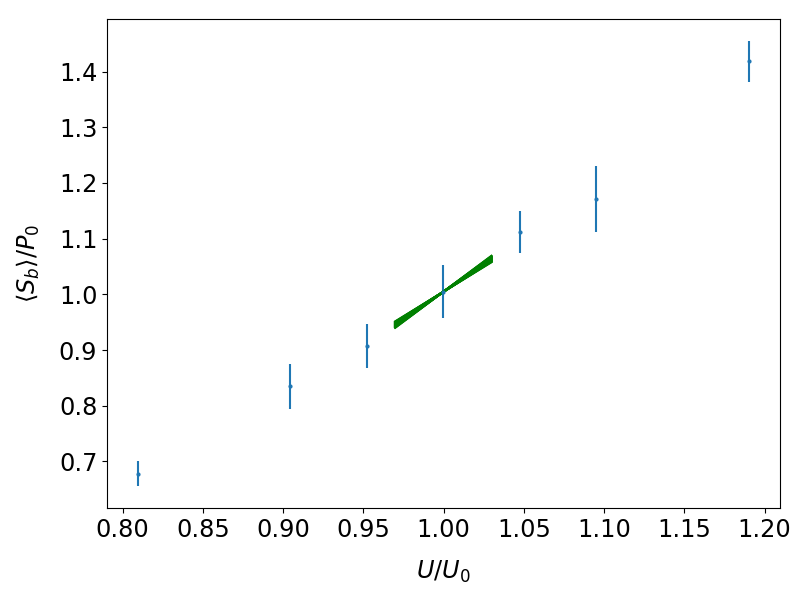}
  \includegraphics[trim=0cm 0cm 0cm 0cm, clip=true, width=0.42\textwidth]{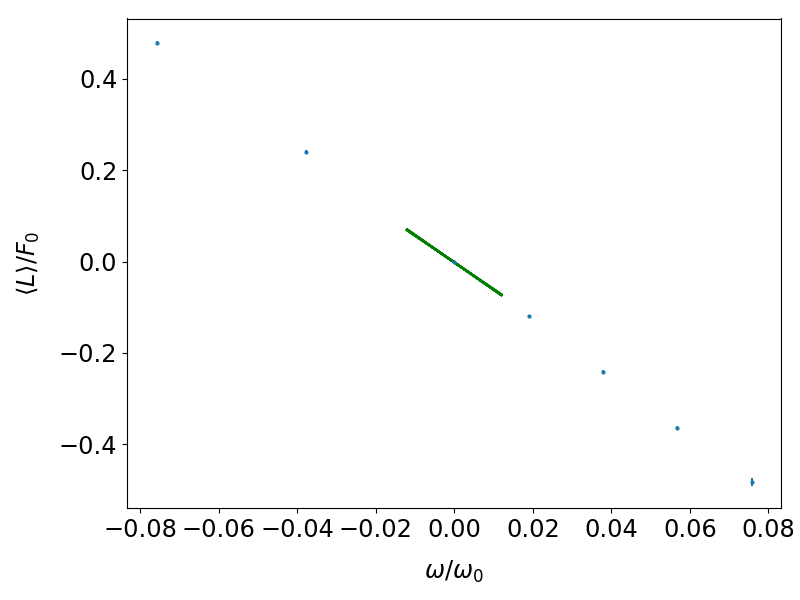}
  \includegraphics[trim=0cm 0cm 0cm 0cm, clip=true, width=0.42\textwidth]{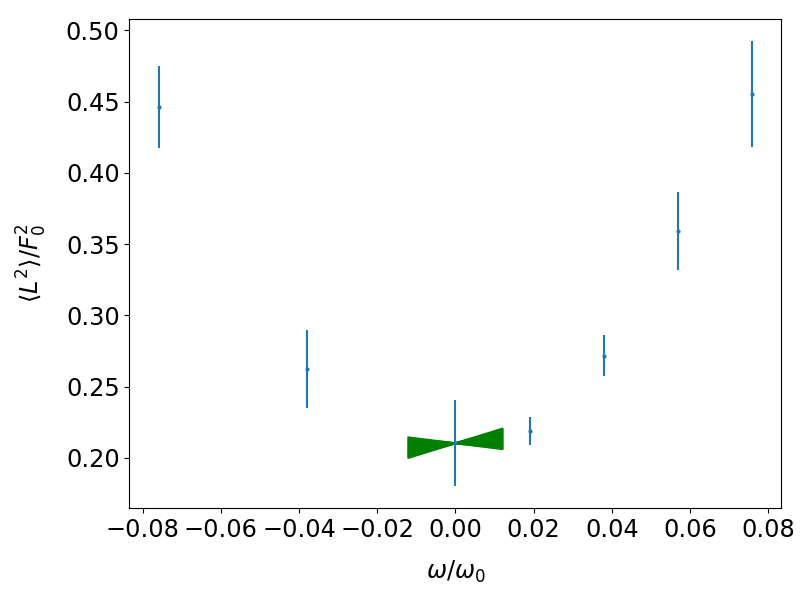}
  \caption{95\% confidence intervals of sensitivities computed by FD-NILSS, indicated by the green wedge.
  Blue vertical bars indicate 95\% confidence intervals of averaged objectives.
  Here all objectives and parameters are normalized.}
  \label{f:result djds}
\end{figure}

Each objective $\avg{J}_\infty$ is approximated by $\avg{J}_{T'}$, which is averaged over $T'=8.7\times 10 ^{-3} = 1148t_0$.
In figure \ref{f:result djds}, we compute each objective with 7 different parameters 
in order to reflect the trend between the parameter and the objective:
this trend will help us validate the sensitivities computed by FD-NILSS.
For the 7 primal simulations, a total number of $6.1\times 10 ^6$ steps of primal solutions are computed.
As we will see later, $T'$ is chosen so that costs of FD-NILSS and primal simulations are similar,
and we will show that the sensitivity computed by FD-NILSS matches the trend suggested by the primal simulation.

To get the uncertainty of averaging objectives over finite time, we divide the history of $J(t)$ into 5 equally long parts.
Denote the objectives averaged over each of the five parts by $J_1,...J_5$.
The corrected sample standard deviation between them are:
\begin{equation}
\sigma' = \sqrt{\frac{1}{4} \sum_{k=1}^{5} (J_k-\avg{J}_{T'})^2} .
\end{equation}
We assume that the standard deviation of $\avg{J}_{T'}$ is proportional to $T'^{-0.5}$. 
Thus, we use $\sigma = \sigma' / \sqrt{5}$ as the standard deviation of $\avg{J}_{T'}$.
We further assume $\pm2\sigma$ yields the 95\% confidence interval for $\avg{J}_{T'}$.
Objectives for different parameters are shown in figure \ref{f:result djds}, where the bars indicate the 95\% confidence intervals.

Each segment in NILSS has 200 time steps, thus the segment length $\Delta T = 1.96\times 10 ^{-6} = 0.259 t_0$.
We set $\epsilon = 10^{-4}$ and the number of segments $K=400$.
Here $\epsilon$ and $\Delta T$ have been checked by the linearity test we discussed in section~\ref{s:remarks},
and $K$ is chosen such that the cost of FD-NILSS is similar to the primal solver.

\begin{figure}[htb]
  \centering
  \includegraphics[trim=0cm 0cm 0cm 0cm, clip=true, width=0.55\textwidth]{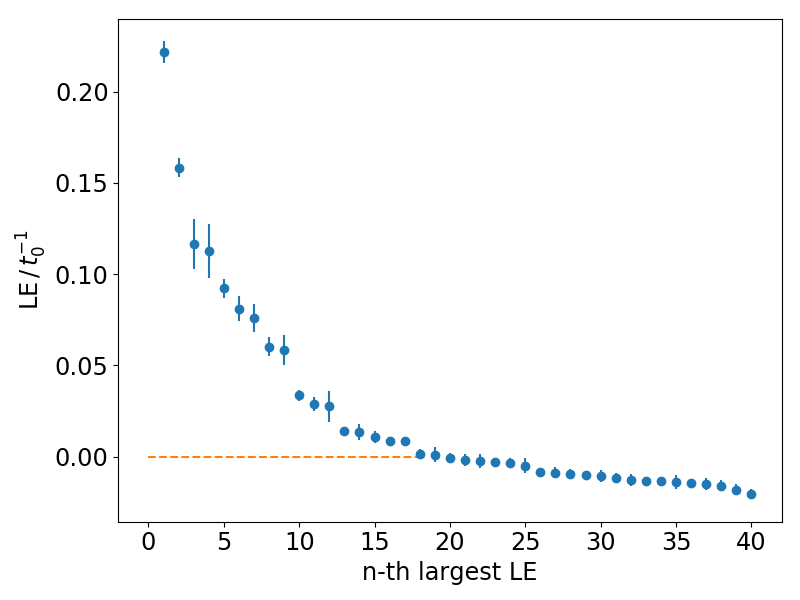}
  \caption{Confidence intervals of the largest 40 Lyapunov exponents (LE), normalized by $t_0^{-1}$.
    The largest LE is 0.22$t_0^{-1}$, meaning in one flow-through time $t_0$, the norm of the first CLV becomes $e^{0.22}=1.25$ times larger.}
  \label{f:LE}
\end{figure}

Our current implementation can not yet inductively add tangent solutions as we discussed in section~\ref{s:remarks}.
Currently, we can only do trial and error to find a large enough $M$, and we selected 40 as our initial guess.
To verify that we have used a large enough $M$,
we use the algorithm given by Benettin \cite{Benettin1980_LE} to find the number of unstable CLVs.
Confidence intervals of LEs are estimated by the smallest interval which bounds the history of an LE and whose size shrinks as $T^{-0.5}$:
this method is the same as in \cite{Ni_NILSS_JCP}.
Figure \ref{f:LE} shows that there are about 17 unstable CLVs, indicating $M$ is large enough.
The LEs, CLVs and shadowing directions of the same physical problem,
on both the current mesh and a finer mesh with twice as many cells,
are studied with more details in \cite{Ni_CLV_cylinder},
which shows that for both meshes, (1) there are only a few unstable CLVs, 
  (2) CLVs are active at different area in the flow field, indicating angles are large between CLVs whose indices are far-apart,
  and (3) shadowing directions exists and can give accurate sensitivities.
Moreover, \cite{Ni_CLV_cylinder} also plots snapshots of CLVs and shadowing directions.

Using above settings, the cost of FD-NILSS is from integrating the primal solution over $400 \times 200 \times 42 = 3.36 \times 10 ^6$ time steps.
Here $K = 400$ is the number of segments, $200$ is the number of time steps in each segment.
$M+2=42$ is the number of primal solutions computed: in the FD-NILSS we need one inhomogeneous tangent and $M = 40$ homogeneous tangents.
Each tangent solution is approximated by the difference between a perturbed trajectory and the same base trajectory: those are 42 primal solutions in total.
The total cost of FD-NILSS is smaller than computing averaged objectives for the 7 parameters in figure~\ref{f:result djds}.
We also remind readers that the marginal cost for a new objective in FD-NILSS is negligible, 
and the marginal cost for a new parameter is about $1/40$ of the total cost.

\begin{figure}[htb]
  \centering
  \includegraphics[trim=0cm 0cm 0cm 0cm, clip=true, width=0.42\textwidth] {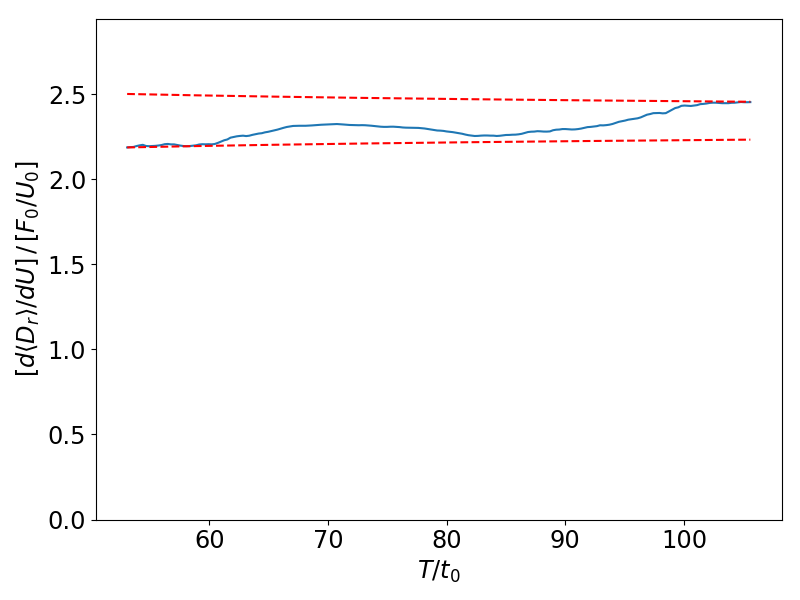}
  \includegraphics[trim=0cm 0cm 0cm 0cm, clip=true, width=0.42\textwidth] {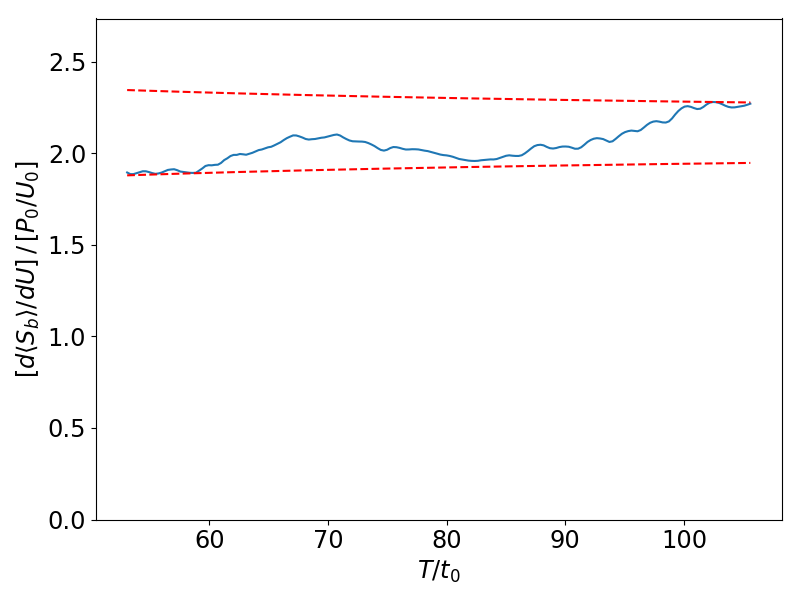}
  \includegraphics[trim=0cm 0cm 0cm 0cm, clip=true, width=0.42\textwidth] {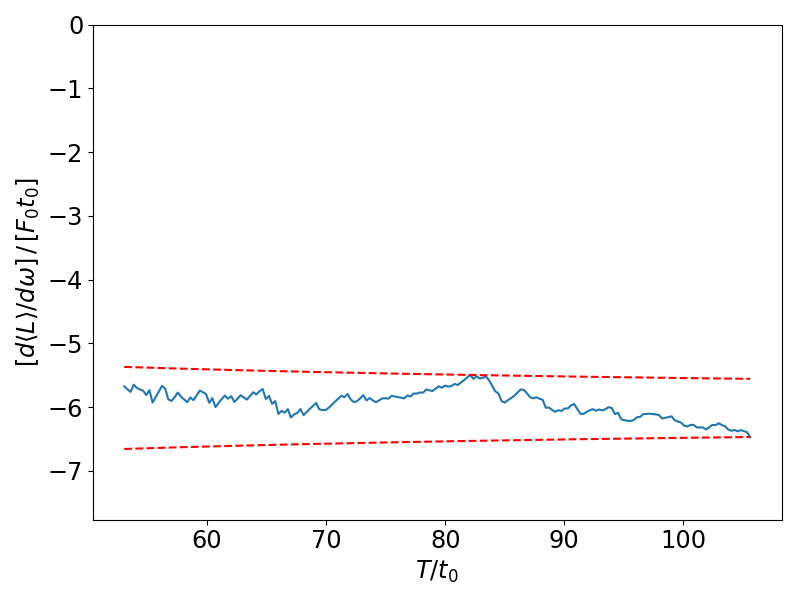}
  \includegraphics[trim=0cm 0cm 0cm 0cm, clip=true, width=0.42\textwidth] {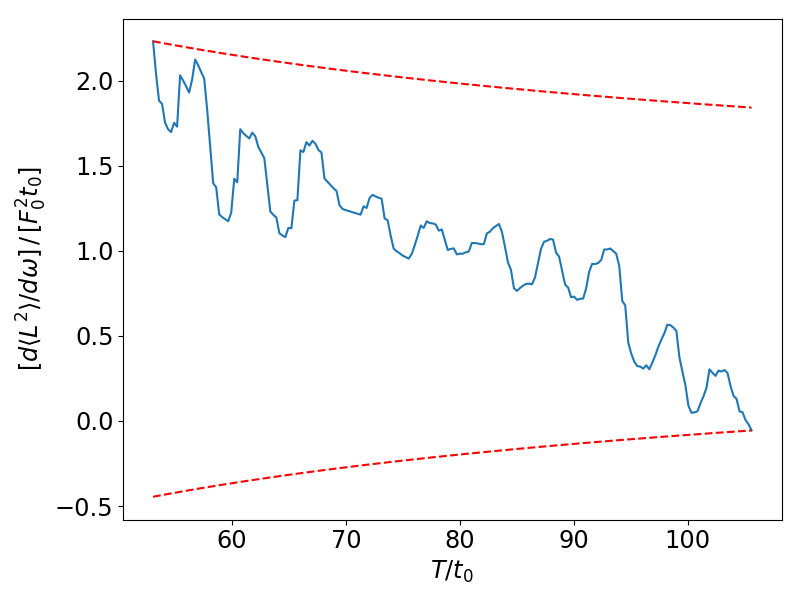}
  \caption{History plots of sensitivities computed by FD-NILSS. All axes are normalized. 
    The dashed lines indicate the smallest encompassing interval whose size shrinks as $T^{-0.5}$.}
  \label{f:djds history}
\end{figure}

The confidence intervals of sensitivities computed by FD-NILSS
are estimated by the smallest interval which bounds the history of the sensitivity and whose size shrinks as $T^{-0.5}$:
this method is given in more detail in \cite{Ni_NILSS_JCP}.
Figure \ref{f:djds history} shows history plots of sensitivities for different pairs of parameter and objective.
In figure \ref{f:result djds} the green wedges are confidence intervals of sensitivities.
Notice that $\avg{L^2}$ attains minimum at $\omega=0$, thus the sensitivity should be almost zero: 
this is why the last plot in figure~\ref{f:djds history} appears not to converge,
since the sensitivity is already very small.

Figure~\ref{f:result djds} validates the sensitivities computed with FD-NILSS, 
since the sensitivities matches the trend between objectives and parameters.
Moreover, the cost of computing sensitivities by FD-NILSS is similar to revealing the trend by evaluating objectives at 7 different parameters.

Another way to compute sensitivities is to perform some function regression among objectives evaluated with different parameters.
However, this regression method requires prescribing a function prototype, the choice of which is typically not obvious.
Even worse, giving confidence intervals to sensitivities computed via regression requires 
prescribing on the space of function prototypes a probability measure,
the choice of which is even less obvious.

\section{Conclusions}

This paper presents the finite difference non-intrusive least squares shadowing (FD-NILSS) algorithm
for computing sensitivities of chaotic dynamical systems.
Unlike NILSS, FD-NILSS does not require tangent solvers, 
and it can be implemented with little modification to existing numerical simulation software.
Numerical results show FD-NILSS can compute accurate sensitivity for the 3-D chaotic flow over a cylinder under Reynolds number 525.
This result also indicates that for real-life engineering problems, FD-NILSS can be an affordable method to compute the sensitivity.

There are several possible future research for the FD-NILSS algorithm.
First, we may investigate the magnitude of the error induced by the finite difference approximation.
We may also investigate if the convergence of the FD-NILSS depends on mesh sizes, time step size, and the finite difference coefficient $\epsilon$.
We can as well experiment different ways of using snapshots to approximate integrations.
For readers who are convinced that FD-NILSS is useful, 
we suggest to further implement NILSS and NILSAS with vectorized linear solvers,
which are faster than FD-NILSS, and could be applied to more chaotic problems with acceptable cost.

\renewcommand*{\bibfont}{\small}
\bibliographystyle{model1-num-names}
\bibliography{MyCollection}

\end{document}